# Observation of Concerted and Stepwise Multiple Dechlorination Reactions of Perchlorethylene in Electron Ionization Mass Spectrometry According to Measured Chlorine Isotope Effects

**Caiming Tang,**[1,2,*] **Jianhua Tan**[3]

[1] State Key Laboratory of Organic Geochemistry, Guangzhou Institute of Geochemistry, Chinese Academy of Sciences, Guangzhou 510640, China

[2] University of Chinese Academy of Sciences, Beijing 100049, China

[3] Guangzhou Quality Supervision and Testing Institute, Guangzhou 510110, China

*Corresponding Author.

Tel: +86-020-85291489; Fax: +86-020-85290009; E-mail: CaimingTang@gig.ac.cn.

**Abstract.** Dechlorinated fragmental ions of organochlorines can be commonly found on electron ionization-mass spectrometry (EI-MS). Yet it is still unclear whether multiple dechlorination reactions of organochlorines taking place in EI-MS are concerted or stepwise. This study investigated the concertedness of the multiple dechlorination reactions of perchlorethylene (PCE) in EI-MS in the light of the observed chlorine isotope effects during different dechlorination reactions along with the detected MS signal intensities. The fluctuant isotope ratios among different ions revealed that the observed multiple dechlorinated reactions in the EI-double focus magnetic-sector high resolution MS (EI-DFS-HRMS) were stepwise. And the changing trends of isotope ratios of the ions detected by EI-quadrupole MS (EI-qMS) might suggest that the observed multiple dechlorination reactions in this MS were concerted. Mechanisms for these concerted/stepwise multiple dechlorination reactions are tentatively interpreted. Besides the evidence of isotope effects, according to the quasi-equilibrium theory, the MS signal intensities also suggested that the multiple dechlorination reactions in the EI-HRMS were likely stepwise, while those in the EI-qMS might be concerted. The distance from the ionization region to the mass analyzer entrance in a mass spectrometer may be a critical factor affecting the dechlorination reactions of ions flying over this distance. This distance in the EI-qMS is significantly shorter than that in the EI-HRMS. Thus, the main detected ions in the EI-qMS might be generated during the ionization and in-source fragmentation processes, while the main ions detected in the EI-HRMS might be produced during the fragmentation of precursor ions flying over the distance. We speculate that the main dechlorination reactions in the ionization region were concerted, while those taking place during the flight of precursor ions over the distance from the ionization region to the mass analyzer entrance were stepwise. This study will shed light on the exploration of dehalogenation processes in analogous reactions and the interpretation of fragmental pathways of halogenated organic compounds in EI-MS.

**Introduction**

A chemical reaction involving multibond breakage or formation may be concerted or stepwise, which has raised concerns for decades [1,2]. The concepts of concerted or stepwise mechanisms are essential to understand the stereochemistry and reactivity of a variety of reactions, such as Diels-Alder reaction, reverse Diels-Alder reaction [3,4], Mclafferty rearrangement [5], Norrish type-I reaction [6], and deiodination reaction [7-10]. A large number of theoretical and experimental studies have dedicated to this issue for revealing the controversial concerted or stepwise mechanisms in various chemical reactions e.g., pericyclic, $S_N2$, E2 aliphatic elimination, and etc [2,6,11-13]. Due to high activation energies of transition states in synchronous multibond processes, multibond reactions are normally non-synchronous [2]. In addition, numerous studies have revealed that many reactions are stepwise rather than concerted. As stated by Kim et al. [6], the concertedness of a reaction should be indicated by that the intermediate lifetime should be within the vibrational period timescale along the reaction coordinate rather than within the rotational period. In this way, the concertedness is examined by referencing the actual nuclear motion to time scales of the reaction coordinate [14,15]. With this criterion, the directly observed concerted reactions are very rare so far [4,15,16].

Multiple dehalogenation reactions with elimination of two or more halogen atoms from a molecule have been studied in terms of concertedness by femtochemistry and quantum chemistry [17,18]. With femtochemistry, Zewail et al. directly observed sequential C-I breakages during the elimination of I atoms in photofragmentation of $CF_2I$-$CF_2I$ [7-10,19]. Lee et al. studied the photodissociation of 1,2-$C_2H_2Br_2$ to Br + Br + $C_2H_2$ using product translational spectroscopy, and concluded that the photodissociation was an asynchronous concerted reaction [20]. Wu et al. reported a concerted elimination of $Br_2^+$ from 1,2-$C_2H_2Br_2$ after laser excitation [21]. Xu et al. studied the ultraviolet photodissociation of bromoform, and found the elimination of Br atoms was sequential [22]. By means of experimental and theoretical femtochemistry, the elimination of Br atoms from 1,3-dibromopropanetime was determined to be stepwise in laser-induced decomposition [23]. Zou et al. investigated the photodissociation of 1,2-dibromo-tetrafluoroethane by photofragment translational spectroscopy with vacuum ultraviolet ionization, and observed that the C-Br bonds dissociation were stepwise [24]. Marvet et al. reported concerted reactions of the elimination of halogen molecules from halogenated alkanes after laser irradiation in femtosecond pump-probe spectroscopy [25].

Elimination of halogen atoms from halogenated organic compounds can usually be observed in electron ionization mass spectrometry (EI-MS). However, the concertedness of multibond-breaking dehalogenation of halogenated organic compounds during fragmentation in EI-MS has never been reported. For symmetric organochlorines, such as perchlorethylene (PCE), the cleavages of multiple C-Cl bonds in EI-MS can be concerted theoretically. Unfortunately, unlike the reactions occurring in femtochemistry, the dechlorination process in EI-MS cannot be directly observed. Therefore, the concertedness of the dechlorination in EI-MS should be judged by other evidences (e.g., isotope effects) rather than reaction timescale that is central to femtochemistry.

Carbon and hydrogen isotope effects have been successfully used to ascertain concertedness or nonconcertedness of some reactions, such as Mclafferty rearrangement and unimolecular ion decomposition [5,26]. In this study, we used chorine isotope effects to investigate whether the dechlorination reactions of PCE in EI-MS were concerted or stepwise. This study will be helpful in the elucidation of various dehalogenation processes and fragmental pathways in EI-MS of halogenated organic compounds.

**Experimental**

*Chemicals and Materials*

Reference standard of PCE (99.0%) was bought from Dr. Ehrenstorfer (Augsburg, Germany). Chromatographic-grade n-hexane was bought from Merck Corp. (Darmstadt, Germany). The standard was accurately weighed and dissolved in n-hexane to prepare a stock solution at 1.0 mg/mL. The stock solution was further diluted with n-hexane to prepare working solutions at 100.0 μg/mL and 1.0 μg/mL. All the standard solutions were stored at -20 ºC condition prior to use.

*Instrumental Analysis*

The working solutions were directly analyzed by gas chromatography-high resolution mass spectrometry (GC-HRMS) or gas chromatography-quadrupole mass spectrometry (GC-qMS). The GC-HRMS system comprised dual gas chromatographers (Trace-GC-Ultra) coupled with a double focus magnetic-sector HRMS and a Triplus auto-sampler (GC-DFS-HRMS, Thermo-Fisher Scientific, Bremen, Germany). The GC-qMS system was an Agilent 7890/5975 GC-MS (Agilent Technologies, Palo Alto, CA, USA).

The columns used in the GC-HRMS and GC-qMS were DB-5MS capillary columns with the same specification (60 m × 0.25 mm, 0.25 μm thickness, J&W Scientific, USA). Detailed descriptions of the temperature programs are provided in the Supporting Information (Table S-1).

The working conditions and parameters of the HRMS are provided as follows: EI source was operated in positive mode; EI energy was 45 eV; ionization source was maintained at 250 ºC; filament current of EI source was 0.8 mA; multiple ion detection (MID) mode was used for data acquisition; dwell time of each isotopologue ion was about 20 ms; mass resolution was ≥ 10000 (5% peak-valley definition) and the HRMS detection accuracy was ± 0.001 u. The HRMS was calibrated in real time with perfluorotributylamine during MID operation.

Structures of PCE and its dechlorinated radical fragments were sketched by ChemDraw (Ultra 7.0, Cambridgesoft), and the exact masses of the molecular and the fragmental isotopologues were calculated with mass accuracy of 0.00001 u. Only chlorine isotopologues were considered. For PCE or a radical fragment containing n Cl atoms, complete isotopologues (n+1) were chosen. The mass-to-charge ratios (*m/z*) of the isotopologue ions were obtained by subtracting the mass of an electron from the exact mass of each isotopologue. The *m/z* values were imported into MID methods for data acquisition. The details related to isotopologues of PCE and the dechlorinated fragments, such as retention times, isotopologue chemical formulas, exact masses and exact *m/z* values are provided in Table S-2.

The working conditions and parameters of the qMS are documented as the following: positive EI source was used; EI energy was set at 45 eV or 70 eV; the temperature of ionization source was kept at 230 °C; selective ion monitoring (SIM) mode was used for data acquisition; dwell time of each isotopologue ion was 30 ms; mass resolution was set at high level (0.2 u). The detailed data including *m/z* values of isotopologue ions for SIM are documented in Table S-2.

*Data Processing*

Chlorine isotope ratio (IR) was calculated as:

$$IR = \frac{\sum_{i=0}^{n} i \cdot I_i}{\sum_{i=0}^{n} (n-i) \cdot I_i} \quad (1)$$

where n is the number of Cl atoms of a molecule or a dechlorinated fragment; i is the number of $^{37}$Cl atoms in an isotopologue ion; $I_i$ is the MS signal intensity of the isotopologue ion i.

All the measured isotope ratios in this study were relative values (raw isotope ratios) without calibration with Standard Mean Ocean Chlorine (SMOC) due to unavailability of the external standard(s) with known Cl isotope composition and structurally identical to PCE. Although the isotope ratios measured in this study were not calibrated to real values referencing to the SMOC, the chlorine isotope fractionation in EI-MS still can be evaluated with these isotope ratios, due to that all the isotopologue ions were detected simultaneously, resulting in accurate relative isotope ratios for the molecular ion and the dechlorinated product ions.

The average MS signal intensity of each isotopologue ion within the whole chromatographic peak was used for isotope ratio calculation. Background subtraction was performed before exporting MS signal intensity by subtracting intensities of the baseline regions adjacent to the corresponding chromatographic peak. Data from six replicated injections were applied to calculate a mean isotope ratio and its standard deviation (1σ).

*Method Performances*

Standard deviations of the measured isotope ratios were within 0.9‰ (Table S-3), which means that precision of the chlorine isotope ratio analysis could fulfill requirements for evaluating Cl isotope effects in EI-MS. Detailed data related to the performances of the developed isotope ratio analysis method are provided in Table S-3.

**Results and Discussion**

*Chlorine Isotope Ratios*

As shown in Figure 1a and Table S-3, the chlorine isotope ratios obtained with GC-HRMS at 45 eV EI energy fluctuated along with the increase of Cl atoms of ions. The product ion with one Cl atom (P-$Cl_1$) showed the lowest isotope ratio (0.28245) among all the ions, and the P-$Cl_2$ exhibited the highest isotope ratio (0.33578). The isotope ratio of the molecular ion was higher than those of the ions P-$Cl_1$ and P-$Cl_3$.

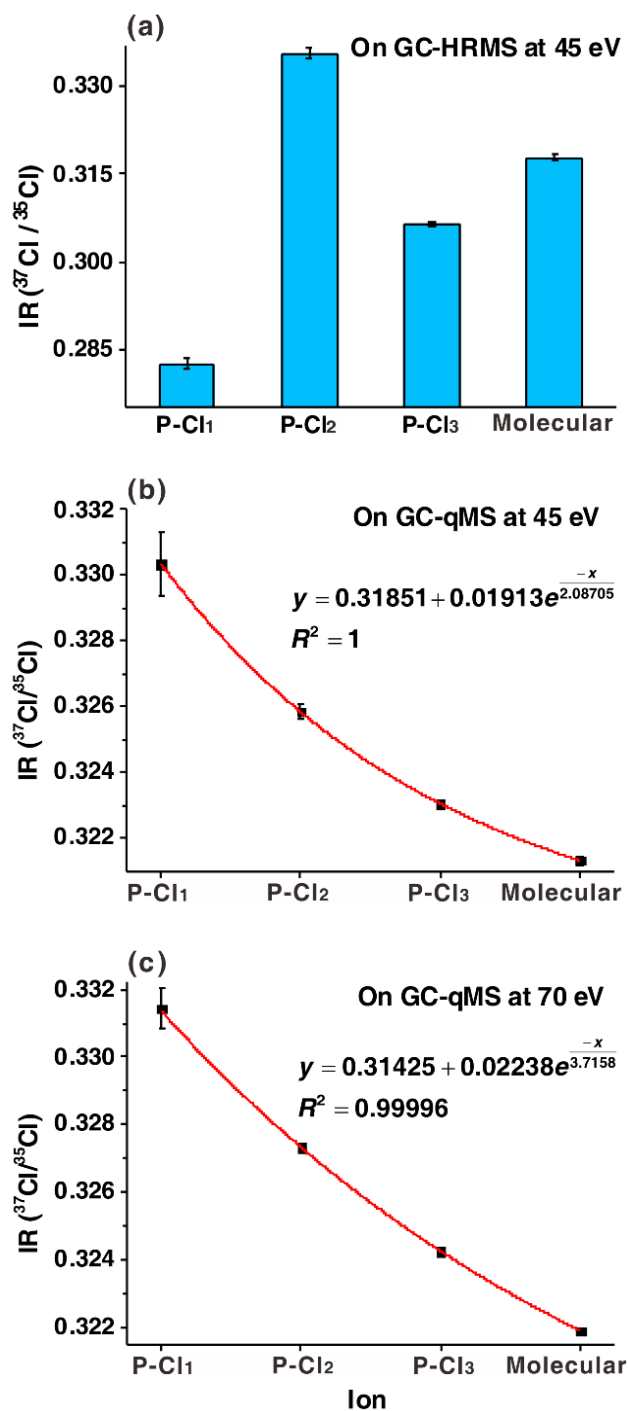

**Figure 1.** Measured chlorine isotope ratios of the molecular and the dechlorinated product ions of perchlorethylene (PCE) by means of gas chromatography-double focus magnetic-sector high resolution mass spectrometry (GC-DFS-HRMS) or gas chromatography-quadrupole mass spectrometry (GC-qMS). **(a)**: isotope ratios obtained with GC-HRMS at EI energy of 45 eV; **(b)**: isotope ratios obtained with GC-qMS at 45 eV; **(c)**: isotope ratios obtained with GC-qMS

at 70 eV; IR: isotope ratio ($^{37}Cl/^{35}Cl$); P-Cl$_n$: dechlorinated product ion possessing n Cl atom(s). The correlations between isotope ratios (y) and the numbers of Cl atoms of ions (x) are fitted with exponential decay functions (ExpDec1 model). Error bars denote the standard deviations (1 σ). The number of injection replicates was six.

The chlorine isotope ratios obtained with GC-qMS are illustrated in Figure 1b-c. With the EI energies of 45 eV and 70 eV, the isotope ratios gradually declined along the order from P-Cl$_1$ to molecular ion. The relationships between the isotope ratios and the number of Cl atoms of ions can be perfectly fitted with exponential decay functions (ExpDec1 model). The functions were

$$y = 0.31851 + 0.01913 e^{\frac{-x}{2.08705}} \qquad (2)$$

and

$$y = 0.31425 + 0.02238 e^{\frac{-x}{3.7158}} \qquad (3)$$

for the relationships obtained with the EI energies of 45 and 70 eV, respectively. At 45 eV, the highest isotope ratio was 0.32687 (P-Cl$_1$), and the lowest was 0.32051 (molecular ion). And at 70 eV, the highest isotope ratio was 0.33145 (P-Cl$_1$), and the lowest was 0.32188 (molecular ion).

*Stepwise Dechlorination Reactions*

The fluctuant isotope ratios among different ions revealed that the observed multiple dechlorinated reactions in EI-HRMS were stepwise. In other words, the reaction pathways were: molecular ion → P-Cl$_3$ → P-Cl$_2$ → P-Cl$_1$. When the molecular ion loses a Cl atom, intramolecular vibrational energy redistribution (IVR) of the available internal energy (AIE) of the product ion takes place. Simultaneously, the bond energies of the remaining C-Cl bonds are changed, due to the structural change from the molecular ion to the product ion. As reported by Zewail et al., the second broken C-I bond had the significantly lower bond dissociation energy (BDE) than the first broken C-I bond. In this study, the breakages of C-Cl bonds might be in the similar scenario. The breakages of the C-Cl bonds of the product ions might require lower energies compared with that of the molecular ion. And the BDEs of the stepwise broken C-Cl bonds were anticipated to be different from each other. As the BDEs and AIEs changed along with the stepwise dechlorination processes, the rate constant $k(E)$ during different dechlorination processes were thus changed according to the quasi-equilibrium theory (QET). According to QET, the rate constant $k(E)$ for a unimolecular reaction is [27,28]:

$$k(E) = \frac{G^*(E-E_0)}{h\rho(E)} \qquad (4)$$

where $E$ is the internal energy of a reaction ion; $E_0$ is the critical energy of the reaction; $G^*(E-E_0)$ is the amount of internal energy states of the transition state of the ion in the energy range from $E_0$ to $E$; $h$ is the Planck's constant, and $\rho(E)$ is the state density of the reaction ion at energy $E$. If an ion lost a Cl atom via two isotopically different pathways with rate constants of $k_l(E)$ and $k_h(E)$ to generate two product ions, the KIE can be expressed as [29]:

$$KIE = \frac{k_l(E)}{k_h(E)} = \frac{G_l^*(E-E_{0l})}{G_h^*(E-E_{0h})} \frac{\rho_h(E)}{\rho_l(E)} \qquad (5)$$

where the subscripts $l$ and $h$ correspond to the light and the heavy isotopes, respectively.

If the reaction positions are structurally identical, reactions are only isotopically different. The KIE can then be expressed as [29]:

$$KIE = \frac{k_l(E)}{k_h(E)} = \frac{G_l^*(E-E_{0l})}{G_h^*(E-E_{0h})} \tag{6}$$

The states density, $\rho(E)$, is the same for the two reaction pathways and thus cannot impact the intra-ion isotope effects.

As can be seen from eq 5 and 6, the KIEs can change when AIEs and critical energies ($E_0$) change. Stepwise multiple dechlorination processes can give rise to intermediates that are dechlorinated radical ions. After electron density redistribution, local structures of the intermediates become different from the corresponding local structures in their parent ions (Note the parent ions are different from the precursor ions here. The parent ions just denote dechlorinating precursor ions, while the precursor ions include both dechlorinating and non-dechlorinating precursor ions). After IVR, the AIEs on the remaining C-Cl bonds are different from those on the corresponding bonds of the parent ions. The changes of local structures and AIEs can alter the critical energies and the internal energies of the remaining C-Cl bonds, leading to changes of KIEs. Thus, KIEs vary in different dechlorination reactions, which may lead to the inconsistent changing tendencies of isotope ratios between different pairs of adjacent ions (Figure 1a).

*Concerted Dechlorination Reactions*

The changing trends of isotope ratios of the ions detected by EI-qMS might show that the observed multiple dechlorination reactions were concerted. The isotope-ratio changing tendencies were substantially different from those observed on EI-HRMS (Figure 1). The changing directions of isotope ratios were consistent from P-Cl$_1$ to molecular ion on EI-qMS, and the isotope ratios exponentially increased with increase of the number of lost Cl atoms.

These observed results may be in line with concerted multiple dechlorination reactions. The pathways of the dechlorination reactions are elucidated in Figure 2.

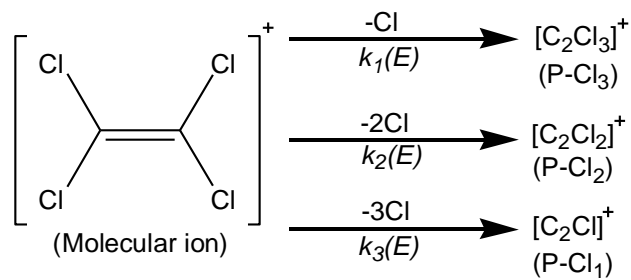

**Figure 2.** Possible dechlorination pathways of PCE on GC-qMS. $k_1(E)$, $k_2(E)$ and $k_3(E)$ are the reaction rate constants of the three dechlorination reactions, respectively.

During a concerted reaction, at least two bonds are broken within a timescale of a vibrational period. Within this timescale, the IVR and electron density redistribution may be far from finish, the internal energies on the reaction bonds and the local structure of the molecule may not change. In other words, for a symmetric compound, all the reaction bonds are structurally and energetically identical. Therefore, according to eq 5 and 6, the KIEs during all the bond breakages are equivalent. Accordingly, the KIE during the dechlorination pathway of molecular ion → P-Cl$_3$ was that in the breaking of one C-Cl bond of the molecular ion($KIE_{C-Cl}$), and the KIE during molecular ion → P-Cl$_2$ was that in the breakages of two C-Cl bonds of the molecular ion [$(KIE_{C-Cl})^2$]; likewise, the KIE in the process of molecular ion → P-Cl$_1$ was that during the breakages of triple C-Cl bonds of the molecular ion[$(KIE_{C-Cl})^3$].

We hypothesize the function of the increment of isotope ratios between two adjacent pathways versus $KIE_{C-Cl}$ as

$$g(KIE_{C-Cl}) = Af(KIE_{C-Cl}) \qquad (7)$$

Then the isotope ratio of the molecular ion (IR$_{mol}$) is

$$IR_{mol} = IR_0 + A[f(KIE_{C-Cl})]^0 = IR_0 + A \qquad (8)$$

where $IR_0$ can be deduced to be the initial isotope ratio of the molecular ion that are going to be dechlorinated (parent ion).

Similarly, the isotope ratios of P-Cl$_3$, P-Cl$_2$ and P-Cl$_1$ can be expressed as

$$IR_{P-Cl_3} = IR_0 + Af(KIE_{C-Cl}) \qquad (9)$$

$$IR_{P-Cl_2} = IR_0 + A[f(KIE_{C-Cl})]^2 \qquad (10)$$

$$IR_{P-Cl_1} = IR_0 + A[f(KIE_{C-Cl})]^3 \qquad (11)$$

respectively.

Generally, the isotope ratio of the product ion with n Cl atom(s) derived from a molecular ion of a polychlorinated organic compound can be expressed as

$$IR_{P-Cl_n} = IR_0 + A[f(KIE_{C-Cl})]^{m-n} \qquad (12)$$

where m is the number of Cl atoms on the molecular ion. This function is very similar to the two fitted functions provided above (eq 2 and 3), indicating the observed multiple dechlorination reactions in EI-qMS might be concerted. Unfortunately, we cannot determine the exact expression of the function $f(KIE_{C-Cl})$.

*Tentative Mechanism Interpretation*

*Perspective from Reaction Energies*. From the perspective of activation energies, multibond reactions are unlikely concerted, due to the activation energies may linearly increase with the number of the breaking bonds [2]. However, the reactions occurring in EI-MS may be far different from reactions taking place in solution. We calculated the summation of the MS signal intensities of all isotopologues of each ion (Figure 3) to explore the concertedness of the multiple dechlorination reactions occurring in EI-MS. As shown in Figure 3a, the generation reactions of P-Cl$_1$ and P-Cl$_2$ in EI-HRMS were unlikely concerted, due to that the signal intensity of P-Cl$_1$ was higher than that of P-Cl$_2$. While the dechlorination reactions in the EI-qMS might be concerted, because the signal intensities of the product ions gradually decreased from P-Cl$_3$ to P-Cl$_1$ (Figure 3b-c).

For parallel fragmentation reactions in EI-MS, the abundance of a molecular ion ($I_M$) can be expressed as follows according to QET [30]:

$$I_M = I_{M_0} e^{-t\sum_{i=1}^{n} k_i(E)} \qquad (13)$$

where $I_{M0}$ is the total abundance of the ionized molecules of PCE; t is the residence time of the ion; n is the number of the parallel fragmentation pathways and $k_i(E)$ is the reaction rate constant of the fragmentation reaction i.

The abundance of a product ion ($I_i$) can be expressed as

$$I_i = \frac{k_i(E) I_{M_0}}{\sum_{i=1}^{n} k_i(E)} \left(1 - e^{-t\sum_{i=1}^{n} k_i(E)}\right) \qquad (14)$$

According to this equation, the relative abundance of a product ion is merely related to the corresponding reaction constant $k_i(E)$.

The rate constant $k_i(E)$ can be expressed as

$$k_i(E) = \frac{G^*(E - E_{0i})}{h\rho_i(E)} \tag{15}$$

where $E_{0i}$ is the critical energy of reaction i; $\rho_i(E)$ the state density of the reaction ion (parent ion) at energy $E$, and can be expressed as

$$\rho_i(E) = \frac{e^N}{\sqrt{2\pi} N^{N-\frac{1}{2}}} \cdot \frac{E^{N-1}}{\prod_1^N \varepsilon_i} \tag{16}$$

where N is the number of weak-coupling classical harmonic oscillators, and $\varepsilon_i$ is the translational energy of the product ion i.

According to eqs 15 and 16, if the internal energy is within a narrow scale, then the reaction constant k(E) is mainly determined by the critical energy $E_0$. As the critical energies roughly linearly increase with the increase of the number of breaking bonds, the product ions produced after more parallel bond breakages likely have lower signal intensities. Therefore, the observed multiple dechlorination reactions in EI-qMS might be parallel (concerted), while those in the EI-HRMS were likely stepwise, particularly the reactions giving rise to the ions P-$Cl_1$ and P-$Cl_2$.

*Perspective from the Differences between the Two Mass Spectrometers*. Since the same type of ionization source (EI+) was used in the GC-HRMS and GC-qMS in this study, the dechlorination reactions happening in both of the sources might be similar. However, the stepwise and the concerted multiple dechlorination reactions were found in EI-HRMS and EI-qMS, respectively. We therefore hypothesize that both stepwise and concerted multiple dechlorination reactions took place in the sources. The distance from the ion source to the dynode of the EI-HRMS is far longer than that of the EI-qMS, which means that the flight trajectory of an ion in the EI-HRMS is far longer than in the EI-qMS. Therefore, the flight time of a detected ion in EI-HRMS is much longer than that in the EI-qMS. Metastable ions can dissociate during the flight from ion source to dynode. It can be anticipated that the dissociation extents of ions in EI-HRMS are larger than those in EI-qMS, due to the longer flight time in EI-HRMS. That is to say, higher ratio of ions can be transformed to metastable ions dissociating in EI-HRMS comparing with that in EI-qMS. However, in this case, the metastable ions dissociating after entering the mass analyzers, i.e., the quadrupole in the qMS and the magnetic sector in the EI-HRMS, cannot affect the detection of the corresponding product ions. In both mass analyzers of qMS and HRMS during a certain dwell time, only the ion with only one $m/z$ can be detected, the product ion(s) generated by a metastable ion cannot be detected during this dwell time, and cannot also be detected in the dwell time(s) set for the product ion(s). Therefore, the metastable-ion dissociation in the mass analyzers cannot contribute to the isotope ratio(s) of the corresponding product ion(s), while can only positively affect (enhance) the isotope ratio of the ion with the same $m/z$ as the metastable ion, showing a normal isotope effect. The ions with more Cl atoms are expected to be more significantly affected by this isotope effect, because they are more likely to be dechlorinated in the mass analyzers.

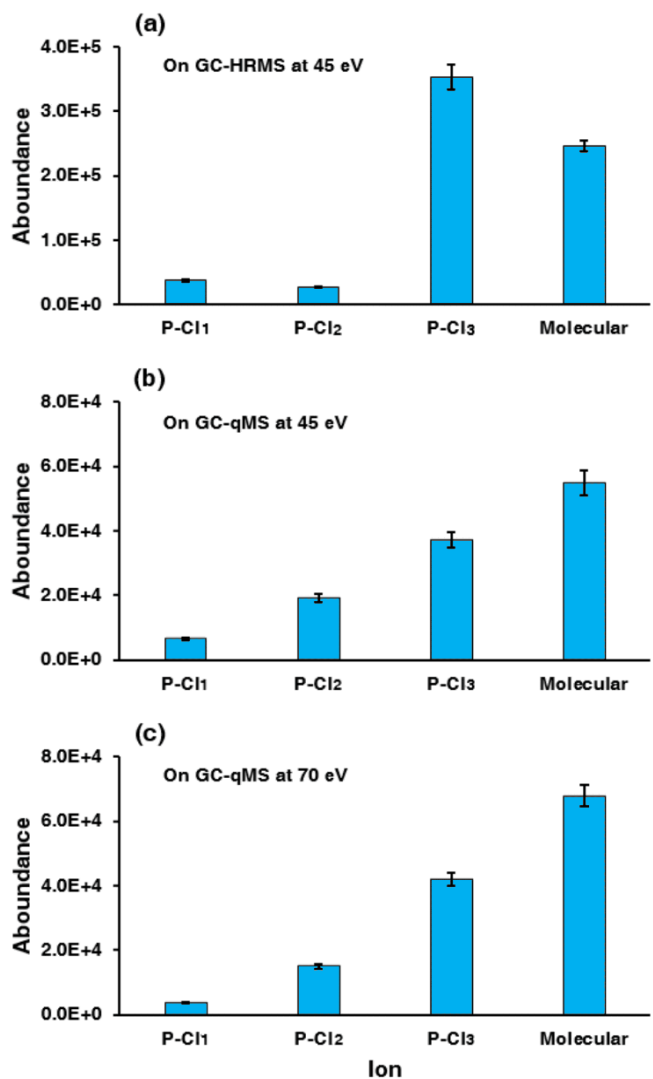

**Figure 3.** Mass spectrometric signal intensities of the molecular and the dechlorinated product ions of PCE on GC-DFS-HRMS or on GC-qMS with different EI energies. **(a)**: signal intensities obtained with GC-HRMS at 45 eV; **(b)**: signal intensities obtained with GC-qMS at 45 eV; **(c)**: signal intensities obtained with GC-qMS at 70 eV.

The distance from ionization region to the entrance of a mass analyzer may be a critical factor impacting the dechlorination processes of ions during the flight over this distance. If the distance is short enough, then most ions cannot dissociate before entering the mass analyzer. Therefore, the measured isotope ratios derive from the ions just generated during the ionization and fragmentation in ion source, provided the isotope effects produced by metastable-ion dissociation in the mass analyzer are not taken into account. If the distance is long enough, then most ions dissociate when overflying this distance, thus the main detected product ions are generated during this dissociation process, provided further dissociation of metastable ions in the mass analyzer is not considered.

The distance from the ion source to the mass analyzer of the EI-qMS is significantly shorter than that of the EI-HRMS. Thus, the main detected ions on EI-qMS may be generated in the ionization and in-source fragmentation, while the main ions detected on the EI-HRMS may be generated during the fragmentation of the ions flying from the ionization region to the entrance of the magnetic sector. We hypothesize that the main dechlorination reactions in the ionization region are concerted, while the main dechlorination reactions during the flight of the ions from the ionization region to the mass analyzer entrance are stepwise. Therefore, the isotope ratios measured with GC-qMS showed the clues to concertedness of the multiple dechlorination reactions, and the changing trends of isotope ratios measured with the GC-HRMS were in line with stepwise processes of the multiple dechlorination reactions.

**Conclusions**

In this study, we applied chorine isotope effects to investigating concertedness of the dechlorination reactions of PCE in EI-MS. The fluctuant isotope ratios of the ions indicated that the multiple dechlorinated reactions in the EI-DFS-HRMS were stepwise. While the isotope ratio changing trends of the ions detected by the EI-qMS suggested that the multiple dechlorination reactions occurring in the EI-qMS might be concerted. Mechanisms for these concerted and stepwise multiple dechlorination reactions were tentatively interpreted. According to the QET, the MS signal intensities showed that the multiple dechlorination reactions in the EI-HRMS were likely stepwise, while those in the EI-qMS might be concerted. The distance from ionization region to the inlet of a mass analyzer in a mass spectrometer might play a critical role in the dechlorination reactions during the flight of ions over this distance. This distance in the EI-qMS is much shorter than that in the EI-HRMS. Thus, the detected ions in the EI-qMS might be mainly generated during the ionization and the in-source fragmentation processes, while those detected in the EI-HRMS may be mainly produced in the fragmentation of the ions flying over the distance. We hypothesize that the major dechlorination reactions taking place in the ionization region were concerted, while those during the flight of ions from the ionization region to the mass analyzer entrance were stepwise. This study will be helpful in the investigation of dehalogenation processes in a variety of reactions and the elucidation of fragmental pathways of halogenated organic compounds in EI-MS.

**Electronic Supplementary Material**

The online version of this article (doi: pending) contains supplementary material, which is available to authorized users.

**Acknowledgements**

We are grateful for Mr. Deyun Liu from Guangzhou Quality Supervision and Testing Institute, China for his assistance in the GC-qMS analysis. This work was partially supported by the National Natural Science Foundation of China (Grant No. 41603092).

**Notes**

The authors declare no competing financial interest.

# Graphical Abstract

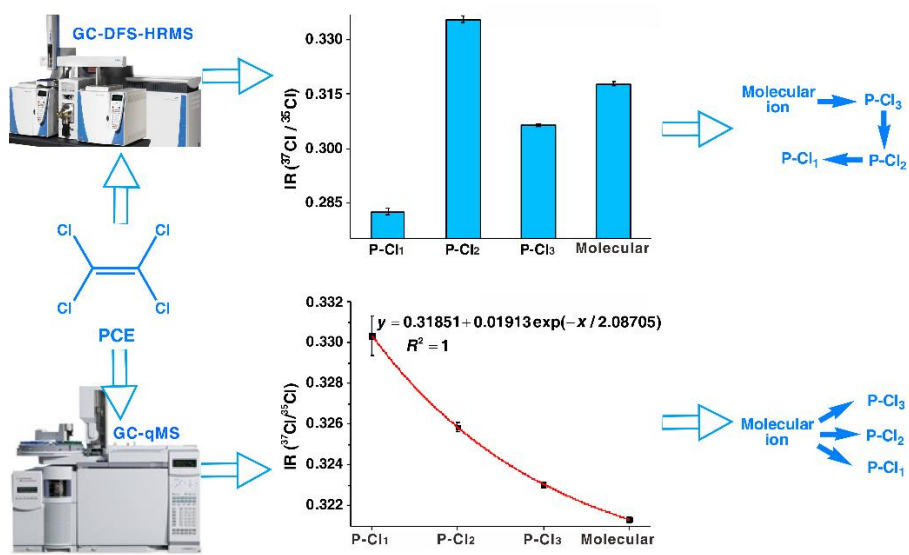